\documentclass[seceq]{ptptex}

\usepackage{graphicx}

\title{
Perturbation theory of multi-plane lens effects 
in terms of mass ratios: 
Approximate expressions of lensed-image positions  
for two lens planes
}

\author{
Koji \textsc{Izumi} and Hideki \textsc{Asada}%
}

\inst{
Faculty of Science and Technology, Hirosaki University, 
Hirosaki 036-8561, Japan
}

\abst{
Continuing work initiated in an earlier publication
(Asada, MNRAS {\bf 394},  (2009) 818), 
we make a systematic attempt to determine, 
as a function of lens and source parameters, 
the positions of images by multi-plane gravitational lenses. 
By extending the previous single-plane work, 
we present a method of Taylor-series expansion 
to solve the multi-plane lens equation 
in terms of mass ratios except for the neighborhood of the caustics. 
The advantage of this method is that it allows 
a systematic iterative analysis and 
clarifies the dependence on lens and source parameters. 
In concordance with the multi-plane lensed-image counting theorem 
that the lower bound on the image number is $2^N$ 
for $N$ planes with a single point mass on each plane,  
our iterative results show how 
$2^N$ images are realized. 
Numerical tests are done to investigate 
if the Taylor expansion method is robust. 
The method with a small mass ratio works well 
for changing a plane separation, 
whereas it breaks down in the inner domain near the caustics. 
}

\begin{document}

\maketitle

\section{Introduction}
Gravitational lensing has become an important subject 
in modern astronomy and cosmology 
\cite{Schneider,Weinberg}. 
It has many applications as gravitational telescopes 
in various fields ranging from extra-solar planets  
to dark matter and dark energy at cosmological scales 
\cite{Refregier}. 
For instance, it is successful in detecting 
extra-solar planetary systems 
\cite{SW,MP,GL,Bond,Beaulieu}.
Gaudi et al.\cite{Gaudi} have found an analogy 
of the Sun-Jupiter-Saturn system through lensing. 
Recently gravitational lensing has been used 
to constrain modified gravity at cosmological scale 
\cite{Reyes}. 

This paper considers the gravitational lensing by 
point-mass systems on multiple planes, 
where the number of planes is arbitrary. 
Such a multi-plane treatment is important. 
In microlensing studies, we usually assume 
a binary lens on a single lens plane. 
In order to discuss its validity, we can 
consider two lens planes and later 
take the limit that two lens planes merge. 
In this way, it will become possible to estimate 
the effect caused by a separation between the double lens planes. 
Another importance is for gravitational lensing in cosmology. 
Clearly, galaxies at different redshifts and dark matter 
inhomogeneities must be described by not a single-plane 
but multi-plane method.  

It has long been a challenging problem to express 
the image positions as functions of lens and source parameters 
\cite{Asada02a,Asada03}.  
For this purpose, we present a method of Taylor-series expansion 
to solve the multi-plane lens equation in terms of mass ratios 
by extending the previous single-plane work 
\cite{Asada09}. 
In particular, we carefully investigate, as a non-trivial task, 
the denominators of the lens equation with singular points. 

The multi-plane lensed-image counting theorem states 
that the lower bound on the image number is $2^N$ for $N$ planes 
with a single point mass on each plane 
(page 458 in Petters, Levine and Wambsganss \cite{PLW} 
and references therein). 
However, the counting theorem tells nothing about 
the image positions. 
Therefore, it is important to discuss {\it how} such image positions 
are realized in an analytical method.

Under three assumptions of 
weak gravitational fields, thin lenses and small deflection angles, 
gravitational lensing is usually described 
as a mapping from the lens plane onto the source plane 
\cite{SW}. 
Bourassa and Kantowski \cite{BKN,BK} introduced a complex notation 
to describe gravitational lensing. 
Their notation was used to 
describe lenses with elliptical or spheroidal symmetry 
\cite{Borgeest,Bray,Schramm}. 

For $N$ point lenses, 
Witt \cite{Witt90} succeeded in recasting the lens equation into 
a single-complex-variable polynomial. 
This is in an elegant form and thus 
has been often used 
in investigations of point-mass lenses. 
The single-variable polynomial due to $N$ point lenses 
on a single plane has the degree of $N^2+1$, 
though the maximum number of images is known as $5(N-1)$ 
\cite{Rhie01,Rhie03,HN06,HN08}.  
This means that unphysical roots are included 
in the polynomial 
(for detailed discussions on the disappearance and appearance 
of images near fold and cusp caustics for general lens systems, 
see also Petters, Levine and Wambsganss \cite{PLW} and references therein).  
Following Asada \cite{Asada09}, we consider the lens equation 
in dual complex variables, so that we can avoid inclusions 
of unphysical roots.

This paper is organized as follows.
In Section 2, the formulation of multi-plane lens systems 
with complex variables is briefly summarized. 
The lens equation is iteratively solved. 
In section 3, we present iterative solutions for a two-plane case 
and give an algorithm for computing image positions 
for an arbitrary number of lens planes in terms of mass ratios. 
In section 4, we discuss how lensed-image positions are realized 
in the present method. 
Section 5 presents numerical tests. 
Section 6 is devoted to the conclusion.

\section{Basic Formulation}
\subsection{Multi-plane lens equation} 
We consider lens effects by N point masses, 
each of which is located at  
different angular diameter distances 
$D_i$ ($i=1, 2, \cdots N$) from the observer, 
where $D_1 \leq D_2 \leq \cdots \leq D_N$. 
For this case, we prepare N lens planes and assume 
the thin-lens approximation for each lens plane 
\cite{BN,YNO}. 
 
All the deflectors line up in small angles and 
all are far away from caustics. 
Note that the above lensing setup is idealized. 
In the real universe, it is rare to find a single isolated lensing 
mass on each plane. 
We consider that contributions from masses at large angles 
are taken into account in the definition of the angular distance 
on average 
\cite{SEF,TAH}. 
In other words, we focus on effects by a local mass distribution 
along the line of sight. 

First of all, angular variables are normalized 
in the unit of the angular radius of the Einstein ring as 
\begin{equation}
\theta_{E}=
\sqrt{\frac{4GM_{tot} D_{1S}}{c^2 D_{1} D_{S}}} , 
\end{equation}
where we put the total mass on the first plane at $D_{1}$, 
$G$ denotes the gravitational constant, 
$c$ means the light speed, 
$M_{tot}$ is defined as the total mass $\sum_{i=1}^N M_i$
and $D_{1}$, $D_{S}$ and $D_{1S}$ 
denote angular diameter distances 
between the observer and the first mass, 
between the observer and the source, and 
between the first mass and the source, respectively.

Recursively one can write down the multi-plane lens equation 
\cite{BN,SEF}. 
In the vectorial notation, 
the two-plane lens equation is written as  
\begin{eqnarray}
\mbox{\boldmath $\beta$}
&=& \mbox{\boldmath $\theta$}-
\left(
\nu_1
\frac{\mbox{\boldmath $\theta$}-\mbox{\boldmath $\ell$}_1}
{|\mbox{\boldmath $\theta$}-\mbox{\boldmath $\ell$}_1|^2} 
\right.
\nonumber\\
 & & ~~~~~~~~
\left. 
+ \nu_2 d_{2}
\frac{\mbox{\boldmath $\theta$}
- \nu_1 \delta_{2}
\displaystyle \frac{\mbox{\boldmath $\theta$}-
\mbox{\boldmath $\ell$}_1}
{|\mbox{\boldmath $\theta$}-
\mbox{\boldmath $\ell$}_1|^2}-
\mbox{\boldmath $\ell$}_2}
{|\mbox{\boldmath $\theta$} 
- \nu_1 \delta_{2}
\displaystyle \frac{\mbox{\boldmath $\theta$}-
\mbox{\boldmath $\ell$}_1}
{|\mbox{\boldmath $\theta$}-
\mbox{\boldmath $\ell$}_1|^2}-
\mbox{\boldmath $\ell$}_2|^2}
\right) , 
\end{eqnarray}
where $\mbox{\boldmath $\beta$}$, $\mbox{\boldmath $\theta$}$, 
$\mbox{\boldmath $\ell$}_1$ and $\mbox{\boldmath $\ell$}_2$ 
denote the positions of the source, image, first and second 
lens objects, respectively.  
Here, $\nu_i$ denotes the mass ratio of each lens object 
as $\nu_i \equiv M_i M_{tot}^{-1}$,  
and we define $d_2$ and $\delta_2$ as  
\begin{eqnarray}
d_2&\equiv&\frac{D_1D_{2s}}{D_2D_{1s}} , 
\\
\delta_2&\equiv&\frac{D_SD_{12}}{D_2D_{1S}} .
\end{eqnarray}

\begin{figure}[t]
\includegraphics[width=14.0cm]{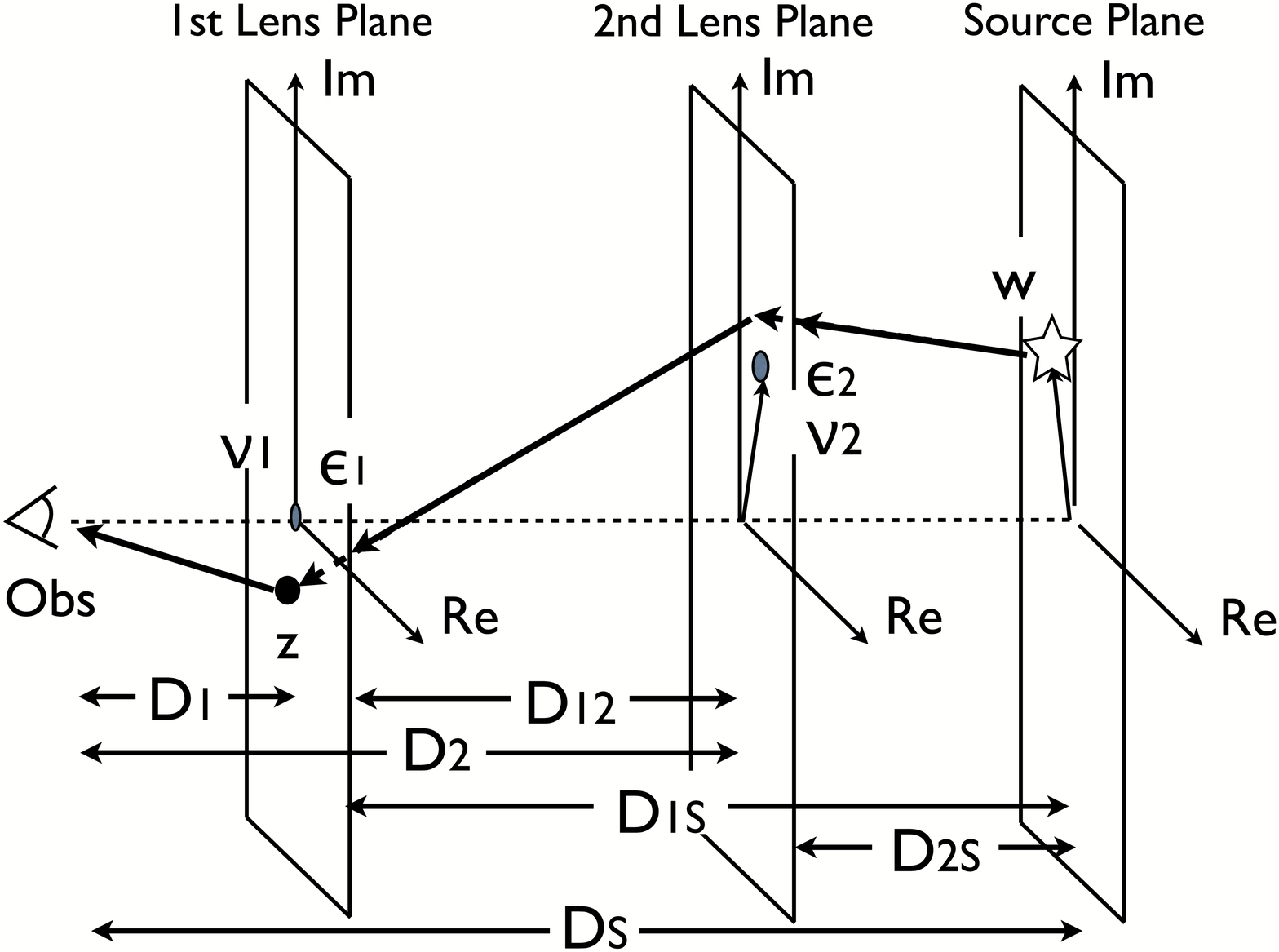}
\caption{
Notation: 
The source and image positions on complex planes are denoted 
by $w$  and $z$, respectively. 
Locations of N masses are denoted by $\epsilon_i$  
for $i=1, \cdots, N$. 
Here, we assume the thin lens approximation for each deflector. 
The angular diameter distances among the observer, source 
and each lens object are also defined. 
}
\label{notation}
\end{figure}

It is convenient to use complex variables when algebraic 
manipulations are done. 
In a formalism based on complex variables, 
two-dimensional vectors for the source, image and lens positions   
are denoted as $w=\beta_x+i \beta_y$, 
$z = \theta_x + i \theta_y$,  
and 
$\epsilon_i=\ell_{ix} + i \ell_{iy}$, 
respectively. 
Figure \ref{notation} shows our notation for 
the multi-plane lens system. 
Here, $z$ is on the complex plane corresponding to 
the first lens object that finally deflects light rays 
and thus $z$ means the direction of a lensed image. 

By employing the complex formalism, the two-plane lens equation 
is rewritten as 
\begin{eqnarray}
w
= z-
\left(
\frac{1-\nu}{z^*}
+ \frac{\nu d_{2}}
{z^* - \epsilon^* -
\displaystyle \frac{(1-\nu) \delta_{2}}{z}}
\right) , 
\label{LensEq}
\end{eqnarray}
where the asterisk $*$ means the complex conjugate 
and we use the identity as $\nu_1 + \nu_2 =1$ to delete $\nu_1$ 
and $\nu$ denotes $\nu_2$.  
Note that we choose the center of the complex coordinate 
as the first mass position. 
Then, we have $\epsilon_1 = 0$ and simply denote 
$\epsilon \equiv \epsilon_2$, which is 
the projected relative position of the second mass 
with respect to the first one. 
The lens equation is non-analytic because 
it contains not only $z$ but also $z^*$.

\subsection{Iterative solutions} 
The mass ratio does not exceed the unity by its definition. 
Therefore, we use a simple method of 
making expansions in terms of the mass ratios. 
One can delete $\nu_1$ by noting the identity as 
$\sum_i \nu_i = 1$. 

Formal solutions are expressed in Taylor series as
\begin{equation}
z=\sum_{p_2=0}^{\infty}\sum_{p_3=0}^{\infty} \cdots
\sum_{p_N=0}^{\infty}
\nu_2^{p_2}\nu_3^{p_3}\cdots \nu_N^{p_N} 
z_{(p_2)(p_3)\cdots (p_N)} , 
\label{formalsolution}
\end{equation}
where the coefficients $z_{(p_2)(p_3)\cdots (p_N)}$ 
are independent of any $\nu_i$.
What we have to do is to determine each coefficient 
$z_{(p_2)(p_3)\cdots (p_N)}$ iteratively. 

At the zeroth order,  
we have always a single-plane lens equation 
as the limit of $\nu_1 \to 1$ ($\nu_2 = \cdots = \nu_N \to 0$). 
We have two roots for it. 
In addition, we have other roots 
for a multi-plane lens equation 
as {\it seeds} 
for our iterative calculations. 
An algorithm for finding these solutions  
is explained in next section.   

Note that the above successive approximation cannot work well 
in the neighborhood of the caustics, 
where the mapping becomes singular. 
Therefore, we focus only on regular regions 
except for the singular domains.

\section{Image Positions}
\subsection{Two lens planes}
At the zeroth order in $\nu$, 
the two-plane lens equation becomes simply 
\begin{eqnarray}
w = z_{(0)} - \frac{1}{z_{(0)}^*} , 
\end{eqnarray}
which is rewritten as 
\begin{equation}
z_{(0)} z_{(0)}^* - 1 = w z_{(0)}^*. 
\label{2-G1}
\end{equation}
The L.H.S. of the last equation is purely real 
so that the R.H.S. must be real. 
Unless $w = 0$, therefore, one can put $z_{(0)}=A w$ 
by introducing a certain real number $A$. 
By substituting $z_{(0)}=A w$ into Eq. ($\ref{2-G1}$), 
one obtains a quadratic equation for $A$ as 
\begin{equation}
w w^* A^2 - w w^* A - 1 = 0 . 
\label{2-G2}
\end{equation} 
This is solved as 
\begin{eqnarray}
A&=& \frac12 
\left(1 \pm \sqrt{1+\frac{4}{w w^*}}\right) 
\nonumber\\
&\equiv&A_{\pm} ,  
\label{2-G3}
\end{eqnarray}
which gives $z_{(0)}$ as $A_{\pm} w$.

In the particular case of $w=0$, Eq. ($\ref{2-G1}$) becomes 
$|z_{(0)}| = 1$, which is nothing but the Einstein ring. 
In the following, we assume a general case of $w \neq 0$. 

Regarding the denominator of Eq. (\ref{LensEq}), 
we make an expansion in $\nu$ as 
\begin{equation}
zz^*-\epsilon^*z-(1-\nu) \delta_2
\equiv \sum_{p=0}^\infty \nu^p f_p , 
\end{equation} 
where we formally obtain 
\begin{eqnarray}
f_0 &=& z_{(0)}z_{(0)}^* - \epsilon^*z_{(0)} - \delta_2 , 
\\
f_1 &=& z_{(0)}z_{(1)}^* + z_{(1)}z_{(0)}^* 
- \epsilon^*z_{(1)} + \delta_2 , 
\\
f_2 &=& z_{(0)}z_{(2)}^* + z_{(1)}z_{(1)}^* 
+ z_{(2)}z_{(0)}^* - \epsilon^*z_{(2)} . 
\end{eqnarray}

By choosing $z_{(0)}$ as $A_{\pm} w$, 
the two-plane lens equation becomes at $O(\nu)$ 
\begin{eqnarray}
z_{(1)} 
+ az_{(1)}^*
= b_1 , 
\end{eqnarray}
where we define 
\begin{eqnarray}
a &\equiv& \frac{1}{(z_{(0)}^*)^2},
\nonumber \\
b_1 &\equiv&-\left(
\frac{1}{z_{(0)}^*}
- d_2  \frac{z_{(0)}}{f_0}
\right) . 
\end{eqnarray}
The above equation is linear in $z_{(1)}$ and thus easily solved as 
\begin{eqnarray}
z_{(1)}=\frac{b_1-ab_1^*}{1-aa^*} . 
\end{eqnarray}

Next, we consider the two-plane lens equation at $O(\nu^2)$. 
It is written as 
\begin{eqnarray} 
z_{(2)} 
+ az_{(2)}^*
&=& 
b_2 , 
\end{eqnarray}
where we define 
\begin{eqnarray}
b_2 &=& 
a z_{(1)}^* 
+ \frac{a (z_{(1)}^*)^2}{z_{(0)}^*}
+ \frac{d_2}{f_0} 
\left( z_{(1)} - z_{(0)} \frac{f_1}{f_0}\right) . 
\end{eqnarray}
This equation is linear in $z_{(2)}$ and thus easily solved as 
\begin{eqnarray}
z_{(2)}=\frac{b_2-ab_2^*}{1-aa^*} . 
\end{eqnarray}

Let us move to $O(\nu^3)$, for which 
the two-plane lens equation is linearized as 
\begin{eqnarray} 
z_{(3)} 
+ az_{(3)}^*
&=& 
b_3 , 
\end{eqnarray}
where we define 
\begin{eqnarray}
b_3 &=& 
-\left[
a \left\{
 a (z_{(1)}^*)^3
 -\frac{2z_{(1)}^*z_{(2)}^*}{z_{(0)}^*}
 - z_{(2)}^*
 + \frac{(z_{(1)}^*)^2}{z_{(0)}^*}
\right\}
\right.
\nonumber \\
& &
~~
\left.
- \frac{d_2}{f_0}
 \left\{ z_{(2)}
  - z_{(1)} \frac{f_1}{f_0} 
  + z_{(0)} 
   \left( 
    - \frac{f_2}{f_0}
    + \frac{f_1^2}{f_0^2} 
   \right)
 \right\}
\right] . 
\end{eqnarray}
This equation is easily solved as 
\begin{eqnarray}
z_{(3)}=\frac{b_3-ab_3^*}{1-aa^*} . 
\end{eqnarray}

In the above, we have considered a rather general case that 
the last term in the two-plane lens equation (\ref{LensEq})
is not divergent. 
Let us investigate the remaining case that 
the denominator of the last term vanishes,   
which is expressed as 
\begin{equation}
z z^* - (1-\nu) \delta_2 = \epsilon^{*} z . 
\end{equation}
The left hand side is real and hence the right hand side 
must be real. 
Therefore, substituting $z = k \epsilon$ for a real number $k$ 
into the above equation leads to a quadratic equation for $k$ as 
\begin{equation}
k^2 - k - \frac{(1-\nu)\delta_2}{\epsilon\epsilon^{*}} = 0 . 
\end{equation}
This is solved to obtain 
$z = z_{+}$ or $z = z_{-}$.  
Here, we define 
\begin{eqnarray}
z_+&\equiv&\frac{\epsilon+\sqrt{\epsilon^2
+4 \nu_1 \delta_2\epsilon(\epsilon^{*})^{-1}}}{2} , 
\\
z_-&\equiv&\frac{\epsilon-\sqrt{\epsilon^2
+4 \nu_1 \delta_2\epsilon(\epsilon^{*})^{-1}}}{2} , 
\end{eqnarray}
where they are not expanded in $\nu$ as an effective renormalization.
The reason for avoiding an expansion in the denominator is as
follows. 
If an expansion in $\nu$ were done in the denominator, 
we would see a third-order pole  (or higher one) in the lens equation. 
This would lead to more complicated iterations. 
In order to avoid it, therefore, we do not 
expand the denominator. 
We exactly treat it.

It should be noted that the zeros of the denominator are the positions 
of the lens objects in a single-plane lens equation \cite{Asada09}, 
whereas the zeros for the present case are 
not the lens positions but located near the lens positions 
with a certain correction due to a separation between the planes. 

We consider the particular case as $z_{(0)} = z_{\pm}$. 
Then, regarding the denominator of Eq. (\ref{LensEq}), 
we make an expansion around $z_\pm$ in $\nu$ as 
\begin{equation}
zz^*-\epsilon^*z-(1-\nu) \delta_2
\equiv \sum_{p=0}^\infty \nu^p g_p , 
\end{equation} 
where we formally obtain 
\begin{eqnarray}
g_0 &=& z_\pm z_\pm^* - \epsilon^* z_\pm - (1-\nu) \delta_2 
\nonumber\\
&=& 0 , 
\label{g-0}
\\
g_1 &=& z_{\pm} z_{(1)}^* + z_{(1)} z_{\pm}^* - \epsilon^* z_{(1)} , 
\\
g_2 &=& z_{\pm} z_{(2)}^* + z_{(1)}z_{(1)}^* 
+ z_{(2)} z_\pm^* - \epsilon^* z_{(2)} . 
\end{eqnarray}
Here, $g_p$ is linear in $z_p$ and $z_p^*$ ($p=1, 2, \cdots$).
Note that a $\nu$-term appears in Eq. (\ref{g-0}), 
because the denominator is exactly treated as a quadratic function. 

At the lowest order in $\nu$, 
the lens equation becomes simply 
\begin{eqnarray}
w = z_{\pm} - \frac{1}{z_\pm^*} - \frac{d_2}{g_1} . 
\end{eqnarray}
This takes the form of $z_{(1)} + a_\pm z_{(1)}^* = b_{\pm 1}$ 
and immediately gives the solution as 
\begin{eqnarray}
z_{\pm (1)}=\frac{b_{\pm 1} - a_\pm b_{\pm 1}^*}{1-a_\pm a_\pm^*} , 
\end{eqnarray}
where we define 
\begin{eqnarray}
a_{\pm} &=& \frac{z_\pm}{z_\pm^* - \epsilon^*} , 
\\
b_{\pm 1} &=& - \frac{d_2z_\pm}
{(z_\pm^* - \epsilon^*)(w-z_{\pm}+\displaystyle\frac{1}{z_{\pm}^*})} . 
\end{eqnarray}

At the next order, 
the lens equation in the complex-conjugated form 
is written as 
\begin{eqnarray}
0 =
z_{\pm(1)} + a z_{\pm(1)}^*+\frac{1}{z_{\pm}^*} 
- \frac{d_2}{g_1} 
\left( 
 z_{\pm(1)} - z_\pm \frac{g_2}{g_1} 
\right) , 
\end{eqnarray}
which is solved as 
\begin{eqnarray}
z_{\pm(2)}=\frac{b_{\pm 2} - a_\pm b_{\pm 2}^*}{1-a_\pm a_\pm^*} . 
\end{eqnarray}
Here, we define 
\begin{eqnarray}
b_{\pm 2}&=&- \frac{1}{z_\pm^* - \epsilon^*}
\nonumber
\\ 
& &
\times
\left( 
(g_1)^2 
\frac{z_{\pm(1)} + a z_{\pm(1)}^*+\displaystyle\frac{1}{z_{\pm}^*} 
- \displaystyle\frac{d_2 z_{\pm(1)}}{g_1}}{d_2z_{\pm}} 
+ z_{\pm(1)}z_{\pm(1)}^*
\right) . 
\nonumber
\\
&&
\end{eqnarray}

Similarly, we find $z_{(3)}$. 

Table $\ref{table1}$ shows a numerical example 
of image positions obtained iteratively and their convergence.

$z_{(1)}$ tells us an order-of-magnitude estimate of 
the effect by a separation between the two lens planes. 
Such a depth effect is characterized by $\delta_2$, 
which enters the iterative expressions through $z_+$ and $z_-$. 

\subsection{Three (or more) planes}
The above procedure for two lens planes 
does not seem to work for an arbitrary number of planes, because 
fifth-order (or higher order) polynomials cannot be solved 
algebraically as shown by Galois 
\cite{Waerden}. 

By iterative procedures, however, one can construct roots 
that are nothing but image positions, 
because we have expansion parameters. 
Let us explain this iterative calculation 
for three-plane lenses as a simple example. 
One can write down the three-plane lens equation. 

First, we neglect $\nu_3$ terms, so that 
the three-plane lens equation can be reduced to 
the two-plane one. 
We have already known how to construct four functions 
denoting image positions for the two-plane lens equation. 
Next, in the similar manner to the two-plane case, 
one can substitute the perturbative image positions 
into the three-plane lens equation. 
Four positions with the correction at $O(\nu_3)$ 
are thus obtained. By using these four linear-order roots, 
one can find four image positions 
at $O(\nu_3^2)$. 
In this way, one can recursively obtain higher order roots.

Other image positions come from the denominator of 
the last term of the three-plane lens equation. 
The denominator takes the same form as the two-plane 
lens equation but with different coefficients 
(obtained by a replacement as $S \to 3$ in the subscripts). 
Hence one can perturbatively construct four other roots. 
By using these four roots as {\it seeds} for further iterations, 
one can construct four roots that can perturbatively 
satisfy the three-plane lens equation. 

Therefore, one can perturbatively construct totally 
$4 + 4 = 8$ image positions. 
Clearly this procedure can be used also for four-plane 
lens systems.

First, we ignore $\nu_4$ terms in the four-plane lens equation, 
so that the equation can be reduced to the three-plane lens equation. 
For $N=3$,  one can find eight image positions as discussed above. 
Hence, one can iteratively obtain an iterative expression 
of eight image positions in terms of $\nu_4$. 
Next, let us take a look at the denominator of the $\nu_4$ term 
in the four-plane lens equation. 
Finding roots of the denominator is essentially 
similar to that of the three-plane lens equation. 
This can be done. 
More eight roots are thus obtained as {\it seeds} 
for iterative calculations. 
One can perturbatively construct totally $8 + 8 = 16$ image
positions as functions of the source and lens parameters. 

We continue the iterative procedure for five (or more) 
lens planes {\it step by step}, so that image positions 
can be perturbatively obtained as functions of the source and lenses. 

Note that two images can merge in the vicinity of the caustics. 
In this paper, we consider only the regular regions, where 
images cannot merge. 
Therefore, zeros of the denominator of the lens equation 
are not degenerate but distinct. 

\begin{table}
\caption{
Example of image positions by 
the two-plane lens. 
We choose $\nu_1=9/10$, $\nu_2=1/10$, $\epsilon=3/2$, 
$w=2$, $D_1/D_S=2/5$, $D_2/D_S=3/5$. 
Iterative results 
(denoted as `0th'. `1st', `2nd' and `3rd') 
show a good convergence for the value 
(denoted as `Num') that 
is obtained by numerically solving the lens equation.   
For the same parameter value, a simple ray-tracing method 
gives numerical values (in the row denoted as 'Ray'). 
}
\begin{center}
    \begin{tabular}{lllll}
\hline
Images & 1 & 2 & 3 & 4
\\
\hline
0th. & 2.414213 & -0.414213 & 1.780776 & -0.280776
\\
1st. & 2.434312 & -0.390217 & 1.731605 & -0.276050
\\
2nd. & 2.430981 & -0.388713 & 1.732327 & -0.275043
\\
3rd. & 2.431474 & -0.388781 & 1.732190 & -0.274861
\\
\hline
Num & 2.431396 & -0.388766 & 1.73220 & -0.274833
\\
\hline
Ray & 2.432 & -0.393 & 1.732 & -0.279 
\\
\hline
    \end{tabular}
  \end{center}

\label{table1}
\end{table}

\section{Realizing Images for N Point Masses} 
Instead of seeking explicit expressions of image positions, 
in this section, 
we discuss how to perturbatively realize 
lensed-image positions 
for arbitrary $N$ planes. 
A hint has appeared in the previous section. 

For $N=2$, the number of the images that are obtained perturbatively 
is four, which equals to $2^N$ for $N=2$. 
By induction, we shall show how at least $2^N$ images are 
realized for $N$ lens planes 
except for the neighborhood of the caustics. 

Let us assume that at least $2^p$ images are realized for $N=p$. 
Note that they are not degenerate, since we do not consider 
the neighborhood of the caustics. 
What we have to do is to show that 
at least $2^{p+1}$ images are realized for $N=p+1$. 

We consider $p+1$ lens planes. 
First, let us ignore the $(p+1)$-th mass term 
in the lens equation, so that 
the equation has the same structure as that for $N=p$. 
By the assumption for $N=p$, therefore, 
the reduced equation with neglecting the $(p+1)$-th mass term 
has at least $2^p$ roots. 

Next, the $(p+1)$-th mass can be considered a new perturber. 
In the lens equation, the denominator of the fraction 
with $\nu_{p+1}$ has $2^p$ zeros.  
Note that it cannot be factored because 
it is a polynomial mixed with $z$ and $z^*$. 
These zeroth-order roots as {\it seeds} lead to 
iterative image positions with the same number. 

In total, {\it at least} $2^p + 2^p = 2^{p+1}$ roots are realized, 
since our iteration method does not exclude additional solutions.  
By induction, we understand how at least $2^N$ images 
are realized for the multi-plane lens equation 
for arbitrary $N$ except for the neighborhood of the caustics. 

The above method of constructing the image positions 
means that the image number inequality $\geq 2^N$ is sharp and 
the lower bound is actually attained. 
Obstruction points at which the backward-traced light ray 
hits a lens object and hence does not reach the source plane 
\cite{PLW} 
play a role in the realization 
in the sense that some of the images are found by investigating 
the neighborhood of obstruction points.

Before closing this section, we make two remarks. 
The first remark is made upon a comparison with numerical methods 
\cite{Hilbert}.  
The present method gives analytical expressions of image positions 
(not their value but their functional forms),  
so that there can be two merits: 
1) Calculations are faster when we obtain the numerical image position. 
2) Dependence on the parameters can be made clearer. 
However, it has a disadvantage, because the result seems 
to take very lengthy expressions. 
The second remark is made on multiple roots. 
The Taylor method assumes that the Taylor series converges, 
whereas at multiple roots (i.e. merging images) it becomes divergent. 
Therefore, it is unlikely that the Taylor method can fix 
the problem when images merge. 
See also the next section.

\section{Numerical tests}  
We perform simple ray-tracing calculations in order to investigate 
if the Taylor expansion method is robust. 
Table \ref{table1} shows that numerical results 
by both methods of the Taylor expansion and the ray tracing 
are in agreement. 
Note that especially the image No. 1 and 3 are 
in good agreement, though the image No. 2 and 4 have 
a few percent difference. 
This is because light rays corresponding to images No. 2 and 4 
pass closer to the primary lens compared with No. 1 and 3 
and therefore numerical errors become relatively large. 

\begin{figure}
\includegraphics[width=84mm]{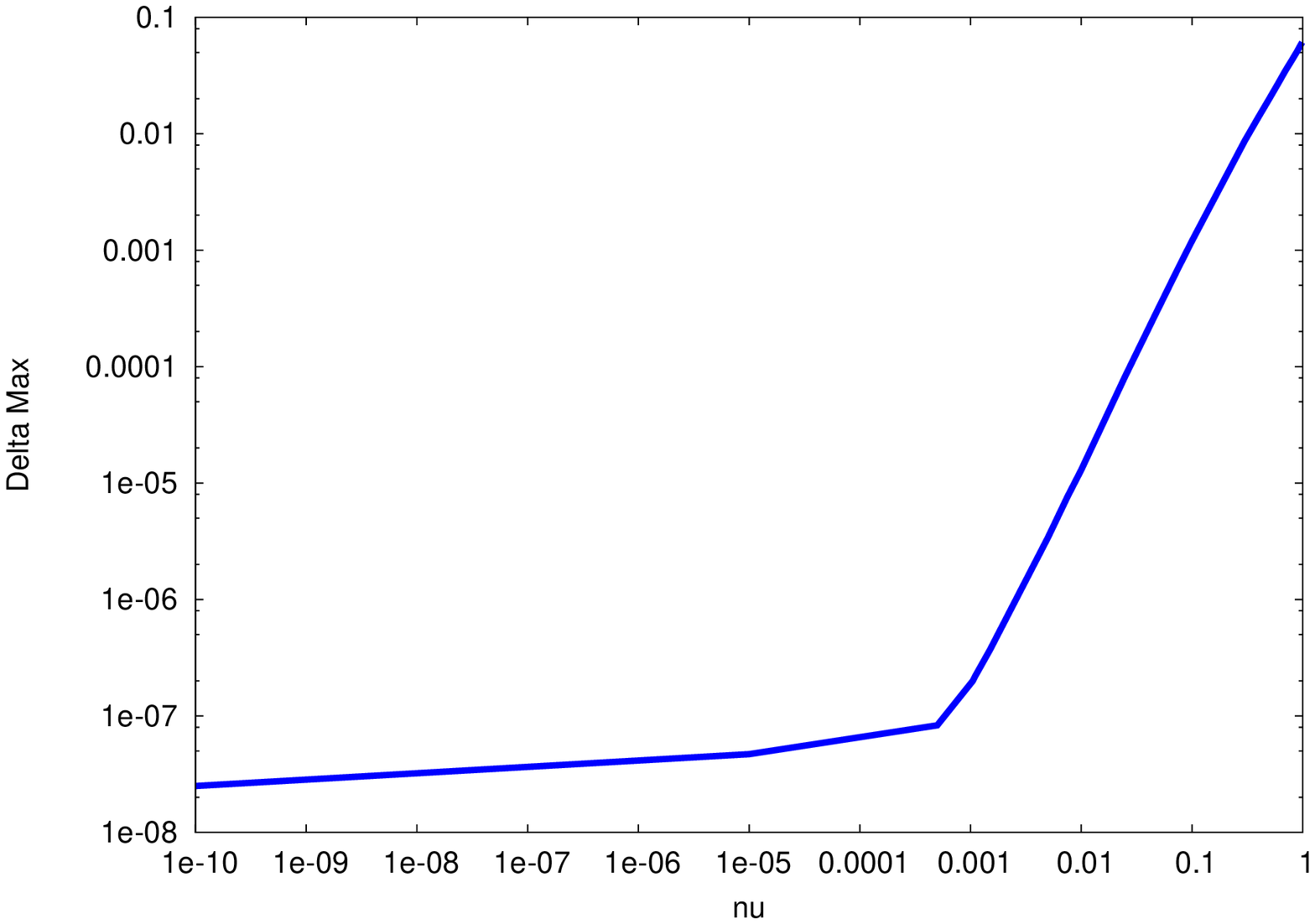}\\
\includegraphics[width=84mm]{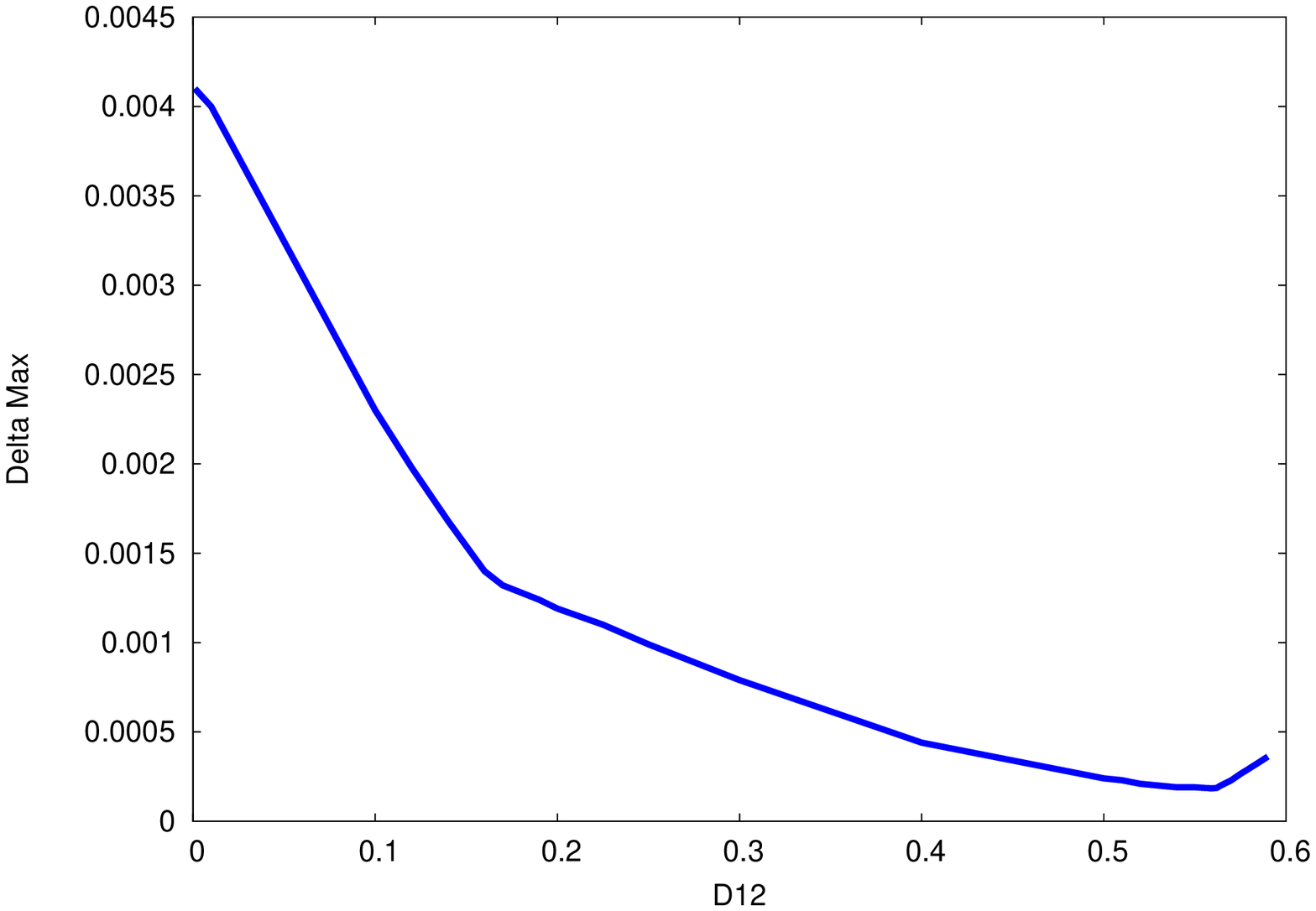}\\
\includegraphics[width=84mm]{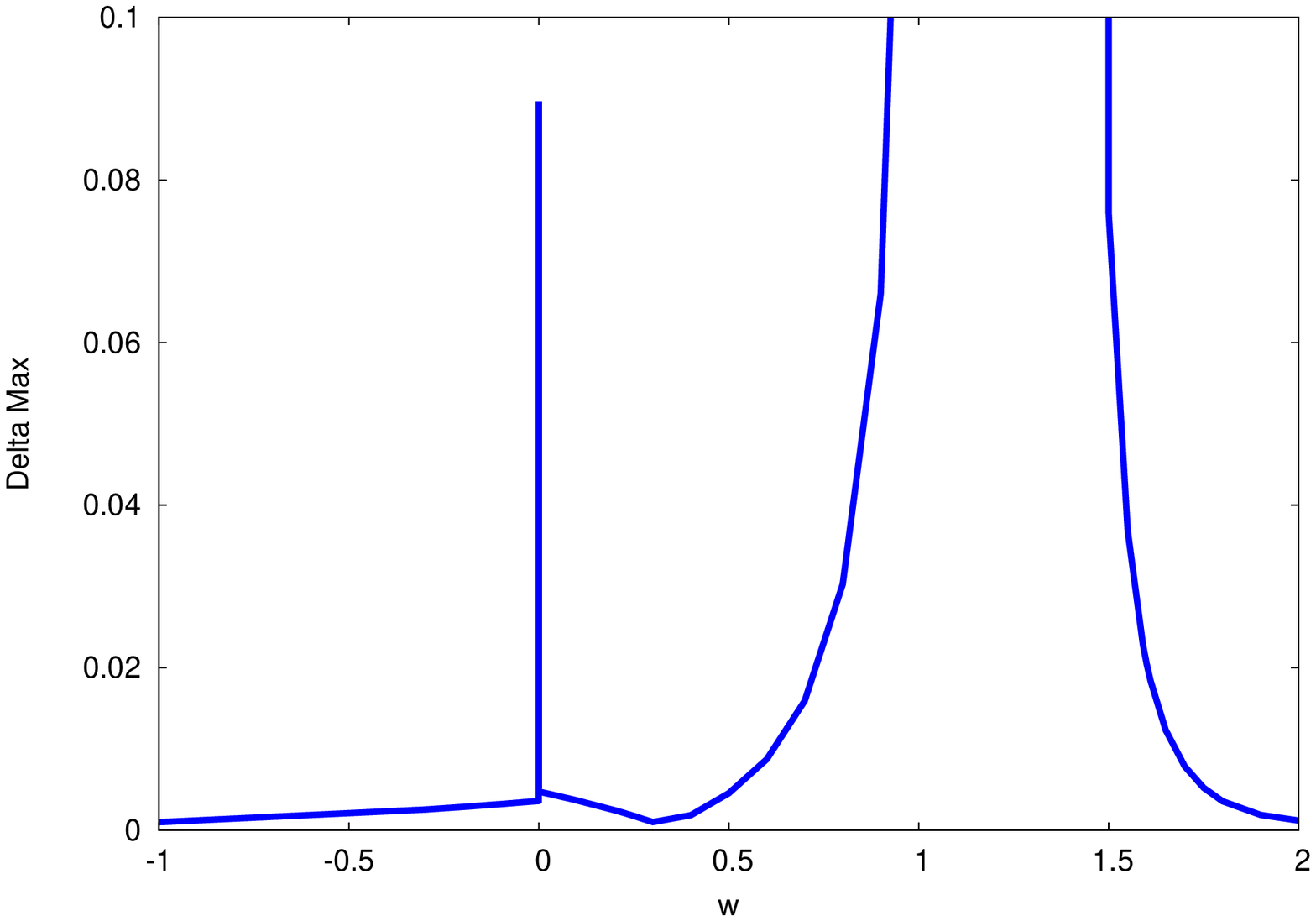}
\caption{
Numerical tests of the accuracy of the Taylor expansion method 
with different parameter values. 
As a reference model for comparisons, 
we choose the model parameters as 
$\nu_1=9/10$, $\nu_2=1/10$, $\epsilon=3/2$, 
$w=2$, $D_1/D_S=2/5$, $D_2/D_S=3/5$, 
which are the same as those in Table 1. 
Top: Mass ratio as $\nu\equiv\nu_2$ is changed. 
Middle: Plane separation as $D_{12}$ is changed. 
Note that $D_{12} < 0.6$ since $D_1 = 0.4$ (normalized by $D_S$). 
Bottom: Source position $w$ is changed along the real axis. 
In actual calculations, 
smaller parameter steps are adopted to investigate the regions 
near the primary and secondary caustics. 
The vertical axis denotes the largest relative error $\Delta_{Max}$ 
of the four images. 
}
\label{accuracy}
\end{figure}

We make also numerical tests for various values 
of the lensing parameters in order to investigate 
the typical size of the cross section, plane separation and mass ratio 
for which the Taylor expansion breaks down. 
Figure \ref{accuracy} shows the accuracy when 
the lensing parameters are numerically changed. 
Here the relative error for each image 
for the chosen parameters 
is denoted as 
\begin{equation}
\Delta \equiv 
\left| \frac{z_{Taylor} - z_{Num}}{z_{Num}} \right| , 
\label{Delta}
\end{equation}
where $z_{Taylor}$ denotes a root obtained by the Taylor expansion 
method (including the third order corrections) and $z_{Num}$ denotes 
the root that is obtained by numerically solving the lens equation. 
$\Delta_{Max}$ denotes the largest error among 
four image positions (for two planes) 
for the chosen parameter values. 
For Figure \ref{accuracy}, we assume the parameter values 
that are the same as those in Table \ref{table1}, 
namely  
$D_1=0.4$, $D_2=0.6$ ($D_{12}=0.2$), 
$\nu=0.1$ (mass ratio), 
$\epsilon=1.5$ (secondary mass position), 
$w=2$ (source position). 
The Taylor expansion method is robust 
for a wide range of the parameters 
as shown by Figure \ref{accuracy}. 
However, it does not work well 
in some cases, 
for instance when a mass ratio is large (i.e. comparable masses), 
or the source is located in the inner region near the caustics. 

In order to quantify the breakdown of the present method, 
we choose the threshold for $\Delta_{Max}$ 
as $0.01$ (one percent). 
The typical size of the cross section  
for which the Taylor expansion method breaks down is approximately 
$\pi \times 0.2^2 \sim 0.1$ (near the primary mass direction) 
and 
$\pi \times 1^2 \sim 3$ (around the secondary one), 
respectively. 
If the threshold is less than one percent, 
the two cross sections are merged as a single one. 
Note that the angle is normalized by the Einstein radius 
with the total mass located at $D_1$. 
The typical size of the mass ratio to invalidate 
the Taylor expansion is 0.3. 
As for the plane separation, the Taylor expansion method 
seems robust.

\section{Conclusion}
We made a systematic attempt to determine, 
as a function of lens and source parameters, 
the positions of images by multi-plane gravitational lenses. 
We presented a method of Taylor-series expansion 
to solve the multi-plane lens equation 
in terms of mass ratios 
except for the neighborhood of the caustics.

In concordance with the multi-plane lensed-image counting theorem 
that the lower bound on the image number is $2^N$ 
for $N$ planes with a single point mass on each plane,  
our iterative results directly show how $2^N$ images 
are realized except for the neighborhood of the caustics.

It is left as a future work to compare the present result 
with state-of-art numerical simulations.

\section*{Acknowledgments}
The authors would like to thank M. Kasai and R. Takahashi 
for stimulating conversations. 
This work was supported in part (H.A.) by a Japanese Grant-in-Aid 
for Scientific Research from the Ministry of Education, 
No. 19035002.

%



\end{document}